\documentclass[11pt]{article}      
\pdfoutput=1
\usepackage{amssymb}
\usepackage{amsmath}
\usepackage{amsthm}
\usepackage[]{graphicx}
\usepackage[]{subfigure}
\usepackage{tensor}
\usepackage{float}
\usepackage{color}
\usepackage{cancel}
\usepackage{setspace}
\usepackage{fancyhdr}
\usepackage{framed}
\usepackage{braket}
\usepackage{tikz}
\usepackage{comment}
\usepackage[noadjust]{cite}

\usepackage[textwidth=6.5in,top=1.2in,bottom=1.2in]{geometry}

\usepackage{setspace}
\usepackage{simplewick}
\usepackage{float}
\usepackage{mathtools}
\usepackage{blkarray, bigstrut} %
\usepackage[bookmarks,linktocpage, colorlinks=true, plainpages = false, citecolor = red,  linkcolor=blue, urlcolor = magenta, filecolor = blue]{hyperref} 
\usepackage{braket}
\usepackage{mathrsfs}
\usepackage[title]{appendix}
\numberwithin{equation}{section}

\begin{document}

\thispagestyle{empty}
\begin{titlepage}
\nopagebreak

\title{\begin{center}\bf Wave packet treatment of neutrino flavor oscillations in various spacetimes\end{center}}

\vfill
\author{P. Sadeghi$^1$\footnote{\href{mailto:psadeghi20@ubishops.ca}{psadeghi20@ubishops.ca}},\, F. Hammad$^{1,2,3}$\footnote{\href{mailto:fhammad@ubishops.ca}{fhammad@ubishops.ca}},\, A. Landry$^3$\footnote{\href{mailto:alexandre.landry.1@umontreal.ca}{alexandre.landry.1@umontreal.ca}},\, T. Martel$^1$\footnote{\href{mailto:tmartel20@ubishops.ca}{tmartel20@ubishops.ca}}}
\date{ }

\maketitle

\begin{center}
	\vspace{-0.7cm}
	{\it  $\,^1$Department of Physics and Astronomy, Bishop's University, 2600 College Street, Sherbrooke, QC, J1M~1Z7
Canada}\\
	{\it  $\,^2$Physics Department, Champlain 
College-Lennoxville, 2580 College Street, Sherbrooke,  
QC, J1M~2K3 Canada}\\
{\it  $\,^3$D\'epartement de Physique, Universit\'e de Montr\'eal,\\
2900 Boulevard \'Edouard-Montpetit,
Montr\'eal, QC, H3T 1J4
Canada}
	\end{center}
\bigskip

\begin{abstract}
We study the effect of gravity on neutrino flavor oscillations when each mass eigenstate is described by a wave packet instead of a plane wave. Two different approaches for implementing the wave packet formalism in the study of neutrino flavor oscillations in curved spacetime are examined. We work with a general static and spherically symmetric spacetime before applying our results to a few spacetime metrics of interest. We first focus on general relativity by examining the effect of the exterior and interior Schwarzschild solutions, as well as the de Sitter-Schwarzschild metric, and then we examine selected metrics from modified gravity models.
\end{abstract}

\end{titlepage}

\setcounter{page}{2}


\section{Introduction}\label{Intro}
The nonzero masses of neutrinos have not only been a bonus for research beyond the Standard Model of particle physics, but also a great boon for theoretical investigations in gravitational physics and cosmology \cite{NeutrinoBook,Review}. Indeed, the weak interactions of neutrinos with matter and their nonzero masses make them a precious tool for probing the gravitational interaction without much contamination from non-gravitational interactions. A key property of neutrinos exploited in such investigations is the well established neutrino flavor oscillations \cite{HistoryNO1,HistoryNO2,HistoryNO3,HistoryNO4,HistoryNO5,HistoryNO6,HistoryNO7,HistoryNO8,HistoryNO9,HistoryNO10}. These oscillations, first proposed by Pontecorvo \cite{Pontecorvo1,Pontecorvo2}, are due to the fact that each neutrino flavor comes as a linear superposition of different mass eigenstates. Since the quantum phase of each mass eigenstate of the superposition is evolving in time at a distinct rate, an initially emitted neutrino flavor might be detected as a different flavor at the end of the neutrino's journey. If, in addition, the neutrino travels inside a gravitational field the time evolution of the quantum phase of each mass eigenstate of the superposition will also be affected differently by the gravitational interaction. One expects then a signature on the flavor oscillations due to classical gravity \cite{Wudka,Ahluwalia1,Fornengo1,Cardall,Fornengo2,Remarks,InsideNO1,NeutrinoInCS,Lambiase1,HamiltonJacobi1,HamiltonJacobi2,InsideNO2,SuperNovaNO,Dvornikov,SpinNO,Geometric,Koutsoumbas,Capolupo}, due to the cosmic expansion of the universe \cite{Koutsoumbas,Xiu-Ju,Xin-Lian,Ren,JunPen,RenPan,Tao,Mandal} and even due to possible quantum gravity effects \cite{Alexandre1,Alexandre2,Marletto}. Furthermore, even discussing the equivalence principle \cite{AcceleratedBlasone} and testing its possible violation is envisaged based on neutrino flavor oscillations \cite{Gasperini,Blasone3}. Because of the ultra-relativistic nature of neutrinos, one often studies the effect of gravity by considering curved spacetimes instead of relying on Newtonian gravity. As such, research on the effect of gravity on neutrino flavor oscillations naturally includes testing theories beyond general relativity as well \cite{Torsion1,Torsion2,Chakraborty,Antonelli,Buoninfante,ConformalNO,ShortReview,Mandal}.

On the other hand, it is well known that because of the intrinsic quantum mechanical uncertainties in the production and detection processes, each mass eigenstate should be treated as a wave packet instead of a plane wave of definite momentum \cite{Giunti4,Paradoxes}. The neutrino flavor transition probability has thus been systematically derived using the wave packet approach in flat spacetime within the framework of quantum mechanics \cite{Keyser,Giunti91,Giunti92,Kiers,Doglov,Giunti97,Giunti02,Akhmedov12}, as well as within the frame work of quantum field theory \cite{Giunti93,Giunti97QFT,Kiers2,Cardall2,Beuthe,Giunti3,Akhmedov,Naumov,Naumov2,Naumov3,WPTreatment2016,DayaBay,CovariantWP1}. The wave packet formalism brings in the concept of coherence length ---first pointed out for neutrino oscillations in Ref.\,\cite{Nussinov}--- which is defined to be the propagation distance over which a quantum system conserves its coherence. The emergence of this length is due to the fact that a wave packet propagates at a specific group velocity. Indeed, if the mass eigenstates of a given neutrino flavor behave as wave packets their different group velocities should cause their separation in space over time. As a result, the initial spatial overlap of the mass eigenstates making the neutrino gradually deceases over time, affecting thus the flavor transition probability itself. In flat spacetime ({\it i.e.}, in the absence of gravity) the difference in the group velocities of a pair of mass eigenstates is proportional to the masses-squared difference of that pair of mass eigenstates. Given the small mass differences observed in neutrinos, the coherence lengths displayed by the latter are larger than the usual distances the detected neutrinos have traveled.

Now, given that in curved spacetimes each momentum of the wave packet evolves differently along the neutrino's path, one expects that gravity would enhance the effect of the wave-packet nature of the mass eigenstates in reshaping the quantum superposition in the initial neutrino flavor. For this reason it is very important to study to what extent neutrino flavor oscillations in curved spacetime are affected when treating each mass eigenstate as a wave packet. Such a study, examined in the appendix of Ref.\,\cite{NeutrinoInCS} based on energy dispersion, has recently been conducted in Ref.\,\cite{ChatelainVolpe} (see also the related work \cite{Petruzziello}) based on momentum dispersion. Our aim in this paper is to extend that study to achieve two additional extremely important goals.

One of our goals is to investigate to what extent can the wave packet formalism bring new insights into the use of neutrino oscillations for testing modified gravity theories. Our second goal is to investigate the two {\it different} possible ways the wave packet formalism in neutrino oscillations could be implemented in curved spacetimes. In fact, the plane wave approach for neutrino oscillations in curved spacetimes relies on computing the Stodolsky integral \cite{Stodolsky} that gives the accumulated quantum phase by each mass eigenstate along the neutrino's path. When moving on to the wave packet approach in curved spacetimes, one then has the option to either (i) first evaluate the Stodolsky integral for each mass eigenstate and then build the wave packet out of these, or (ii) first build the wave packet and then find out the effect of the curved spacetime on the latter as it propagates from the production point to the detection point. The first approach was adopted in Ref.\,\cite{ChatelainVolpe} for a Schwarzschild spacetime. However, the second approach has its own merits and advantages for, as mentioned above, the necessity for the wave packet formalism arises because of the quantum uncertainties inducing momentum dispersion in {\it both} the production and detection processes. Building the wave packet right from the neutrinos source is what takes this fact into account. More important, however, is the fact that the second approach allows one, in addition, to account for the effect of gravity on the spatial overlap of the mass eigenstates. 

We shall thus systematically examine in this paper both approaches and compare their results by working with the most general static and spherically symmetric metric, and then apply our results to specifically chosen metrics of interest. We shall first examine the exterior and interior Schwarzschild solutions as they are useful for describing the exterior and interior gravitational fields of non-rotating astrophysical objects. Still within general relativity, and given the observed cosmic expansion of our Universe, we also apply our results to the case of the de Sitter-Schwarzschild spacetime representing the gravitational field created by a massive object inside an expanding universe. We then examine different metrics that emerge as solutions to specific modified gravity theories already examined in Refs.\,\cite{Chakraborty,Buoninfante} using the plane wave approach to neutrino oscillations. 

The outline of this paper is as follows. In Sec.\,\ref{sec:FlatWP}, we recall the wave packet treatment of neutrino oscillations in flat spacetime by outlining the main steps for computing the flavor transition probability and justifying our choice of the tools we are going to use in curved spacetime. In Sec.\;\ref{sec:StodolskyFirst}, we derive the general expression of the transition probability for a general static and spherically symmetric spacetime by evaluating first the accumulated phase of each mass eigenstate with the Stodolsky integral before building the wave packet. In Sec.\;\ref{sec:WavePacketFirst}, we derive the general expression of the flavor transition probability by building the wave packet first before computing its accumulated phase at the detection point. In Secs.\;\ref{sec:ExteriorSchw}, \ref{sec:InteriorSchw} and \;\ref{sec:deSitterSchw} we examine the effect of, respectively, the exterior Schwarzschild solution, the interior Schwarzschild solution and the de Sitter-Schwarzschild spacetime. In Sec.\;\ref{sec:ModifiedGravity}, we examine two different static and spherically symmetric spacetime metrics emerging, respectively, from the $\mathcal{R}^2$-gravity model and from an infinite-derivative modified gravity model. In the last section, we conclude this paper with a brief summary and discussion.

\section{The neutrino wave packet in flat spacetime}\label{sec:FlatWP}
The flavor neutrino $\ket{\nu_\alpha(x,t)}$ at any spacetime point $(x,t)$\footnote{For simplicity and clarity, we consider throughout this paper a one-dimensional propagation of the neutrinos. Indeed, the one-dimensional case renders the interpretation of our results and the physical meaning of each term of the cumbersome expressions derived here easily extractable without losing generality.} is expressed as a linear superposition of the mass eigenstates $\ket{\nu_j(x,t)}=\psi_j(x,t)\ket{\nu_j}$ via the Pontecorvo-Maki-Nakagawa-Sakata (PMNS) unitary mixing matrix $U_{\alpha j}$ as follows \cite{NeutrinoBook}:
\begin{equation}\label{Ket}
\ket{\nu_\alpha(x,t)}=\sum_jU^*_{\alpha j}\,\psi_j(x,t)\ket{\nu_j}.
\end{equation}
The first letters $\alpha,\beta$ of the Greek alphabet will be used in this paper to denote a neutrino flavor, whereas the Greek letters $\mu,\nu$ will be used to denote spacetime coordinate indices. The Latin letters $j,k$ will denote mass eigenstates of masses $m_j$ and $m_k$, respectively. Thus, $\psi_j(x,t)$ denotes the coordinate space wavefunction corresponding to the mass eigenstate $\ket{\nu_j}$ that belongs to the orthonormal Fock states, $\braket{\nu_j|\nu_k}=\delta_{jk}$. In the literature, two distinct methods are usually adopted for computing the probability for an $\alpha$ neutrino produced at position $x_A$ to be detected as a $\beta$ neutrino at position $x_B$ at time $T$. The first method relies on the flavor transition amplitude, which is found by projecting the state $\ket{\nu_\alpha(x,T)}$ onto the state $\ket{\nu_\beta(x-L)}$ representing the detected neutrino at the coordinate distance $L$ (see, e.g., Ref.\,\cite{Giunti97}):
\begin{equation}\label{ProbabilityWithStates}
    \mathcal{P}(\nu_\alpha\rightarrow\nu_\beta)=\int{\rm d}T\,\bigg|\!\int{\rm d}x\,\braket{\nu_\beta(x-L)|\nu_\alpha(x,T)}\bigg|^2.
\end{equation}
The spatial integral over $x$ gives the flavor transition amplitude, whereas the integral over $T$ leads to the average probability because the time variable is not usually measured in neutrino oscillation experiments. The neutrinos source and the laboratory detectors typically operate over times much longer than
the oscillation times\footnote{We set throughout the paper $\hbar=c=1$.} $\sim E/\Delta m_{jk}^2$, where $E$ is the neutrino energy and $\Delta m_{jk}^2=m_j^2-m_k^2$.

The second method consists in relying on the density matrix operators $\hat{\rho}_\alpha(x)$ and $\hat{\mathcal{O}}_\beta(x-L)$ associated, respectively, with the production and detection processes. The operator $\hat{\rho}_\alpha(x)$ should be obtained by averaging again $\hat{\rho}_\alpha(x,T)$ over the unmeasured time $T$. The time-dependent density matrix operator $\hat{\rho}_\alpha(x,T)$ is computed from the state (\ref{Ket}) by $\hat{\rho}_\alpha(x,T)=\ket{\nu_\alpha(x,T)}\bra{\nu_\alpha(x,T)}$. The density matrix operator $\hat{\mathcal{O}}_\beta(x-L)$ associated with the detection process at the coordinate distance $L$ is obtained from $\hat{\rho}(x)$ by the replacement of $x$ by $x-L$. The flavor transition probability is then given in the matrix density method by (see, e.g., Ref.\,\cite{Giunti4}),
\begin{equation}\label{DMProbability}
\mathcal{P}(\alpha\rightarrow\beta)={\rm Tr}\left[\hat{\rho}_\alpha(x)\hat{\mathcal{O}}_\beta(x-L)\right]=\int {\rm d}x\sum_j\bra{\nu_j}\hat{\rho}_\alpha(x)\hat{\mathcal{O}}_\beta(x-L)\ket{\nu_j}.
\end{equation}
The two methods are equivalent, but we will adopt the method (\ref{ProbabilityWithStates}) as it makes our investigation based on wave packets in curved spacetimes more transparent and less cumbersome. 

In the wave packet study of neutrino oscillations, one usually considers for each mass eigenstate a Gaussian distribution of momenta centered on the mean momentum $\bar{p}_i$ with a dispersion width $\sigma_{pP}$ associated with the production process. In flat spacetime, the coordinate space wavefunction $\psi_j(x,t)$ of each neutrino mass eigenstate $\ket{\nu_j}$ of energy $E_j(p)$ and of three-momentum $p_j$ is then given at any spacetime point $(x,t)$ by
\begin{equation}\label{FlatWavePacket}
\psi_j(x,t)=\int_{-\infty}^{+\infty}\frac{{\rm d}p_j}{2\pi}\left(\frac{2\pi}{\sigma_{pP}^2}\right)^{\frac{1}{4}} e^{-\frac{(p_j-\bar{p}_j)^2}{4\sigma_{pP}^2}}e^{-iE_j(p)t+ip_jx}.
\end{equation}
This normalized Gaussian integral thus simply represents a weighted superposition of plane waves, having each the phase $-E_j(p)t+p_jx$ at a given spacetime point $(x,t)$.

In order to evaluate the integral (\ref{FlatWavePacket}), one needs to express the energy $E_j(p)$ in terms of the momentum $p_j$. For that purpose, one notices that the Gaussian function $\exp[-\frac{(p_j-\bar{p}_j)^2}{4\sigma_{pP}^2}]$ is rapidly suppressed away from the mean momentum $\bar{p}_j$. Therefore, one may, to a good approximation, just expand $E_j(p)$ in terms of $p_j$ around $\bar{p}_j$ to a few powers of the difference $p_j-\bar{p}_j$. However, in order to obtain the term responsible for the widening over time of the wave packet, one needs to expand $E_j(p)$ at least up to the second power of $p_j-\bar{p}_j$ as follows:
\begin{align}\label{FlatEExpansion}
E_j(p)&=\bar{E}_j+\frac{\partial E_j(p)}{\partial p}\bigg|_{p=\bar{p}_j}\left(p_j-\bar{p}_j\right)+\frac{1}{2!}\frac{\partial^2 E_j(p)}{\partial p^2}\bigg|_{p_j=\bar{p}_j}\left(p_j-\bar{p}_j\right)^2\nonumber\\
&\equiv\bar{E}_j+v_j\left(p_j-\bar{p}_j\right)+\gamma_j\left(p_j-\bar{p}_j\right)^2.
\end{align} 
Here, $\bar{E}_j$ represents the average energy and $v_j$ represents the group velocity of the wave packet associated to the mass eigenstate $\ket{\nu_j}$. Recalling that for relativistic neutrinos we have the mass-shell relation $E_j(p)=(p_j^2+m_j^2)^{1/2}$, and neglecting terms of order $(m_j^2/\bar{p}_j^2)^2$ and higher, we easily find $v_j\approx\,1-m_j^2/2\bar{p}_j^2$ and
    $\gamma_j\approx m_j^2/2\bar{p}_j^3$. Substituting these into Eq.\,(\ref{FlatEExpansion}), and then plugging the latter into the Gaussian integral (\ref{FlatWavePacket}), we arrive at the following expression for the wave packet:
\begin{equation}\label{FinalFlatWP}
    \psi_j(x,t)=\frac{(2\sigma_{pP}^2/\pi)^{\frac{1}{4}}}{\sqrt{1+4i\sigma_{pP}^2\gamma_jt}}\exp\left[-i\bar{E}_jt+i\bar{p}_jx-\frac{\sigma_{pP}^2(x-v_jt)^2}{1+4i\sigma_{pP}^2\gamma_jt}\right].
\end{equation}
The time-dependent factor multiplying the exponential in this expression is responsible for the decrease over time of the wave packet's amplitude. The first two terms inside the exponential are responsible for the oscillating wave imprisoned under the bell-shaped envelop. The last term inside the exponential is responsible for the propagation of this envelop of the wave packet as well as for the time variation of the shape of the wave packet. The latter becomes indeed wider over time because of the time-dependent denominator in that term. This widening of the envelop restores some of the spatial overlap ({\it i.e.}, the coherence) lost due to the difference in group velocities of the mass eigenstates. Note that if we did not go up to the order $(p_j-\bar{p}_j)^2$ in the expansion (\ref{FlatEExpansion}), we would not have had the factor $\gamma_j$, for which case we easily see from expression (\ref{FinalFlatWP}) that neither the wave packet's amplitude nor the width of the latter would change over time. 

Since Eq.\,(\ref{ProbabilityWithStates}) requires an integration over time, to describe the mass eigenstate $\ket{\nu_j(x,T)}$ at the detection point we will approximate the time variable in the denominators in Eq.\,(\ref{FinalFlatWP}) by setting $T\approx x_B-x_A=L$ \cite{WPTreatment2016}. This is justified in Minkowski spacetime by the ultra-relativistic nature of the neutrinos and by the fact that the error this approximation would bring in is already beyond $\mathcal{O}[(\Delta m_{jk}^2/E^2)^2]$. In fact, for a minimal uncertainty wave packet we have $\sigma_p\sigma_x=\frac{1}{2}$, where $\sigma_x$ is the spatial width of the wave packet. On the other hand, flavor oscillations vanish for $\sigma_x\gtrsim L_{\rm osc}$, where $L_{\rm osc}$ is the oscillation length of the wave inside the envelop of the wave packet. Since $L_{\rm osc}\sim E/\Delta m_{jk}^2$, we conclude that the condition for oscillations is $\sigma_p\lesssim\Delta m_{jk}^2/E$ \cite{Beuthe}. As such, the contribution of the denominator is non-negligible only for very large $x$, which we take to be at $x\sim L$, {\it i.e.}, at the detection point.  To describe the detected neutrino $\ket{\nu_\beta(x-L)}$ using the wave packet (\ref{FinalFlatWP}), we only need to set $t=0$, and replace $x$ by $x-L$ and replace the production momentum uncertainty $\sigma_{pP}$ by the detection momentum uncertainty $\sigma_{pD}$ in the expression \,(\ref{FinalFlatWP}).  

Implementing these prescriptions in the wave packet (\ref{FinalFlatWP}), the flavor transition amplitude of Eq.\,(\ref{ProbabilityWithStates}) takes the following form:
\begin{equation}\label{ComputedFlatAmplitude}
    \int{\rm d}x\braket{\nu_\beta(x-L)|\nu_\alpha(x,T)}=\sum_{j}U^*_{\alpha j}U_{\beta j}\sqrt{\frac{2\tilde{\sigma}_{pj}^2}{\sigma_{pP}\sigma_{pD}}}\exp\left[-i\bar{E}_jT+i\bar{p}_jL-\tilde{\sigma}_{pj}^2(L-v_jT)^2\right],
\end{equation}
where,
\begin{equation}\label{TildeSigma}
    \frac{1}{\tilde{\sigma}_{pj}^2}=\frac{1}{\sigma_{pP}^2}+\frac{1}{\sigma_{pD}^2}+4i\gamma_j L.
\end{equation}
We introduced, for convenience, the ``modified dispersion width" $\tilde{\sigma}_{pj}$ which is specific to each mass eigenstate as it carries the information about the widening of the wave packet thanks to the coefficient $\gamma_j$. In this form, which to the best of our knowledge has not been previously used in the literature, the effect of the wave packet's widening due to gravity will be very clear when we move on to curved spacetimes. Note, also, that the integration over $x$ to arrive at the amplitude (\ref{ComputedFlatAmplitude}) is necessary to take into account the momentum dispersion behind the wave-packet nature of the detected neutrino caused by the detection process. If we wish to ignore the uncertainties due to the detection process, we should compute, instead, the amplitude $\braket{\nu_\beta|\nu_\alpha(L,T)}$ in which the produced neutrino would still be represented by a wave packet $\ket{\nu_\alpha(L,T)}$ at distance $L$ from its emission, but the detected neutrino would simply be a Fock state $\ket{\nu_\beta}$. In this case, the amplitude would be given again by Eq.\,(\ref{ComputedFlatAmplitude}) after simply removing $\sigma_{pD}$ from the denominator inside the square root and removing $1/\sigma_{pD}^2$ from the right-hand side of Eq.\,(\ref{TildeSigma}). 

Using the approximation $\bar{p}_j\approx\bar{E}-m_j^2/2\bar{E}$, where we take $\bar{E}$ to be the average energy common to all the mass eigenstates, and then squaring the magnitude of the amplitude (\ref{ComputedFlatAmplitude}) and integrating the result over the time $T$, we find the flavor transition probability to be given by
\begin{equation}\label{ComputedFlatProbability}
\mathcal{P}(\alpha\rightarrow\beta)\propto\sum_{j,k}U^*_{\alpha j}U_{\beta j}U_{\alpha k}U^*_{\beta k}\exp\left[-2\pi i\left(\frac{L}{L_{\rm osc}}+\frac{1}{4\pi}\tan^{-1}\left[\frac{(\gamma_j-\gamma_k)L}{\sigma_x^2}\right]\right)-\frac{L^2}{\mathcal{Z}_{\rm coh}^2}\right].
\end{equation}
Here, we have set,
\begin{equation}\label{FlatLengths}
    L_{\rm osc}=\frac{4\pi\bar{E}}{|\Delta m_{jk}^2|},\qquad \mathcal{Z}_{\rm coh}=\frac{\sqrt{\tilde{\sigma}_{pj}^2v_j^2+\tilde{\sigma}_{pk}^{*2}v_k^2}}{\tilde{\sigma}_{pj}\tilde{\sigma}_{pk}^*|v_j-v_k|},
\end{equation}
and we have introduced, for convenience, the width $\sigma_x$ of the wave packet in coordinate space, such that  $\sigma_x^2=\sigma_{xP}^2+\sigma_{xD}^2=\frac{1}{4}\sigma_{pP}^{-2}+\frac{1}{4}\sigma_{pD}^{-2}$. The first term inside the exponential represents the effect of the traveled distance $L$ by the wave packet on the flavor transition probability. This effect is quantified by the ratio of the effective traveled distance to the natural oscillation length $L_{\rm osc}$ of the neutrinos. The second and third terms inside the exponential carry the effect of the widening of the wave packets which restores back some coherence, and the damping effect caused by the spatial separation over time of the mass eigenstates' wave packets. This combination emerges thanks to the complex nature of the quantity $\mathcal{Z}_{\rm coh}$. In fact, using that $v_j\approx 1-m_j^2/2\bar{E}^2$ and $\gamma_j\approx m_j^2/2\bar{E}^3$, the real and imaginary parts in $\mathcal{Z}_{\rm coh}$ can be separated to the leading order in $\Delta m_{jk}^2/\bar{E}^2$ as follows:
\begin{equation}\label{ZSplit}
\mathcal{Z}_{\rm coh}=\frac{4\sqrt{2}\bar{E}^2\sigma_x}{|\Delta m_{jk}^2|}\left(1-\frac{m_j^2+m_k^2}{2\bar{E}^2}\right)+i\frac{\sqrt{2}L}{2\sigma_x\bar{E}}.
\end{equation}
The real term on the right-hand side of Eq.\,(\ref{ZSplit}) is just the usual coherence length $L_{\rm coh}$ in flat spacetime obtained up to first-order correction in $m_j^2/\bar{E}^2$. The imaginary term in Eq.\,(\ref{ZSplit}) is what is responsible (because of the dispersion of the wave packet) for {\it not fully} restoring the lost coherence due to the different group velocities of the wave packets. To see this, we plug expression (\ref{ZSplit}) into the exponential in Eq.\,(\ref{ComputedFlatProbability}) so that the latter takes the form,
\begin{equation}\label{ClarifiedFlatProbability}
\mathcal{P}(\alpha\rightarrow\beta)\propto\sum_{j,k}U^*_{\alpha j}U_{\beta j}U_{\alpha k}U^*_{\beta k}\exp\left[-2\pi i\left(\frac{L}{L_{\rm osc}}\!+\!\frac{1}{4\pi}\tan^{-1}\left[\frac{2\pi L}{\sigma_x^2\bar{E}^2L_{\rm osc}}\right]\!-\!\frac{\sqrt{2}}{2\pi\sigma_x \bar{E}}\frac{L^3}{L^3_{\rm coh}}\right)-\frac{L^2}{L_{\rm coh}^2}\right].
\end{equation}
The coherence length appears now also inside the oscillating term of the flavor changing probability, but it appears inside a term that comes with a negative sign. Therefore, the transition phase, which consists of the content inside the parentheses, becomes increased due to the widening of the wave packet but only partially because of the dispersion in the latter. In the next sections we will adapt this procedure and these tools to the case of curved spacetimes. 
\section{The neutrino wave packet in curved spacetime}
\subsection{Evaluating the Stodolsky phase first}\label{sec:StodolskyFirst}
In curved spacetimes, the four-momentum $p^\mu$ of each of the mass eigenstates is spacetime-dependent. As a consequence, when a quantum particle of four-momentum $p^\mu$ is {\it propagating} between spacetime points $(x_A,0)$ and $(x_B,t)$ in a curved spacetime of metric $g_{\mu\nu}$, one needs to evaluate the quantum phase accumulated by the particle at the arrival point as given by the Stodolsky integral, $\Phi_{AB}=\int_A^B g_{\mu\nu}p^\mu{\rm d}x^\nu$ \cite{Stodolsky}. Relying on the quantum phase in the investigation of the effects of gravity on quantum systems has been an active research approach since very early on\footnote{See, e.g., Refs.\,\cite{Papini1965,Anandan1,Anandan2,COWReview,Audretsch,Cai1989,Cai1990,QPhaseI,QPhaseIII,QPhaseII,Nandi2009,Josephson,COWBall} for a few selected works and the references therein.}. Therefore, if the detected neutrinos at $(x_B,T)$ are to be described by a wave packet, one way of finding the curved-space analog of the flat-space Gaussian integral (\ref{FlatWavePacket}) would be to write,
\begin{equation}\label{CurvedWavePacket}
\psi_j(x_B,x_A;T)=\int_{-\infty}^{+\infty}\frac{{\rm d}p_j}{2\pi}\left(\frac{2\pi}{\sigma_{pP}^2}\right)^{\frac{1}{4}} e^{-\frac{(p_j-\bar{p}_j)^2}{4\sigma_{pP}^{2}}}e^{i\Phi_{AB}}.
\end{equation}
The final accumulated phase $\Phi_{AB}$ of each plane wave is thus computed before building the wave packet. Therefore, the mean momentum $\bar{p}_j$ in this integral should be taken to have its value at the detection point $(x_B,T)$ but the Gaussian dispersion width $\sigma_{pP}$ is the one introduced at the production point. This is the approach adopted in Ref.\,\cite{ChatelainVolpe}. In this section, we will also adopt this approach to extract the flavor transition probability and the coherence length for a more general metric than the Schwarzschild metric examined in Ref.\,\cite{ChatelainVolpe}.

The first thing we need to define then is the metric $g_{\mu\nu}$ for the background spacetime. Let the latter be static and spherically symmetric, described by a metric of the form,
\begin{equation}\label{GeneralMetric}
    {\rm d}s^2=-\mathcal{A}(r){\rm d}t^2+\mathcal{B}(r){\rm d}r^2+r^2\left({\rm d}\theta^2+\sin^2\theta{\rm d}\phi^2\right).
\end{equation}
Here, the arbitrary functions $\mathcal{A}(r)$ and $\mathcal{B}(r)$ depend only on the radial coordinate $r$. For simplicity, we assume a radial propagation of neutrinos throughout the rest of this paper. Within such a spacetime, a mass eigenstate of mass $m_j$ and of four-momentum $p_j^\mu=m_j\frac{{\rm d}x^\mu}{{\rm d}s}$ of a radially propagating neutrino gives rise to the following energy $E_j(p)$ and momentum $p_j$ as perceived by an observer at infinity:
\begin{equation}\label{InfinityObservables}
    E_j(p)=-g_{00}p_j^0=\mathcal{A}(r)m_j\frac{{\rm d} t}{{\rm d} s},\qquad p_j=g_{rr}p_j^r=\mathcal{B}(r)m_j\frac{{\rm d}r}{{\rm d}s}.
\end{equation}
The spacetime (\ref{GeneralMetric}) possesses a timelike Killing vector $K^\nu=(1,0,0,0)$ as well a spacelike Killing vector $R^\mu=(0,0,0,1)$. The first vector implies the conserved energy $-K_\mu m_j\frac{{\rm d}x^\mu}{{\rm d}s}=E_j(p)$, while the second vector implies the conservation of the angular momentum which need not concern us here as we deal with radially propagating neutrinos. Using the mass-shell condition, $-m_j^2=g_{\mu\nu}p_j^\mu p_j^\nu$, we find
\begin{equation}\label{massshell}
-m_j^2=-\frac{E_j^2(p)}{\mathcal{A}(r)}+m_j^2\mathcal{B}(r)\left(\frac{{\rm d}r}{{\rm d}s}\right)^2.
\end{equation}
Isolating ${\rm d}t/{\rm d}s$ in terms of $E_j(p)$ from the first identity in Eq.\,(\ref{InfinityObservables}) and then dividing the result by the ${\rm d}r/{\rm d}s$ that we isolate from this mass-shell condition (\ref{massshell}), we find the following link between the coordinates $t$ and $r$: 
\begin{equation}\label{dr/dt}
\frac{{\rm d}t}{{\rm d}r}=\sqrt{\frac{\mathcal{B}(r)}{\mathcal{A}(r)}}\left(1-\frac{m_j^2\mathcal{A}(r)}{E_j^2(p)}\right)^{-\frac{1}{2}}.
\end{equation}
On the other hand, plugging the second identity in Eq.\,(\ref{InfinityObservables}) into the result (\ref{massshell}), we arrive at,
\begin{equation}\label{pinE(p)}
p_j=E_j(p)\sqrt{\frac{\mathcal{B}(r)}{\mathcal{A}(r)}}\left(1-\frac{m_j^2\mathcal{A}(r)}{E^2_j(p)}\right)^{\frac{1}{2}}.
\end{equation}
Now, in order to evaluate the phase $\Phi_{AB}$ in Eq.\,(\ref{CurvedWavePacket}) one might be tempted to evaluate the Stodolsky integral by using the link (\ref{dr/dt}) between the time and space coordinates. Doing so, one would compute the accumulated phase $\Phi_{AB}$ by writing, $\int_A^B(-E_j(p){\rm d}t+p_j{\rm d}r)=\int_A^B(-E_j(p)\frac{{\rm d}t}{{\rm d}r}+p_j){\rm d}r$. The reason why this method of computing the Stodolsky phase does not work in the wave-packet treatment of neutrino oscillations is that one must integrate over $t$ and $r$ {\it independently}, as done in Ref.\,\cite{ChatelainVolpe}. This is indeed what offers the possibility to integrate over the unmeasured final time $T$. Note, however, that in doing so one discards the link (\ref{dr/dt}) between the two variables even though such a link is itself derived from the mass-shell condition (\ref{massshell}). The latter condition is supposed to be satisfied (in the quantum mechanical framework we are working here) by each mass eigenstate when the latter are viewed as propagating and by any quantum particle to which the Stodolsky phase integral could be prescribed. Therefore, letting the mass eigenstates propagate all the way to the detector before using them to build a wave packet, as we do here, and yet not allowing them to obey the condition (\ref{dr/dt}) is another issue that justifies considering the second approach that will be dealt with in Sec.\,\ref{sec:WavePacketFirst}. 

Therefore, we compute the accumulated phase $\Phi_{AB}$ by integrating now independently over time and space. Using again the expression (\ref{pinE(p)}) of $p_j$ in terms $E_j(p)$ after expanding it up to the first order in $m_j^2/E_j^2(p)$, we find
\begin{align}\label{WrongStodolskyPhase}
\Phi_{AB}&\approx-E_j(p)\int_A^B{\rm d}t+E_j(p)\int_A^B\sqrt{\frac{\mathcal{B}(r)}{\mathcal{A}(r)}}\,{\rm d}r-\frac{m_j^2}{2E_j(p)}\int_A^B\sqrt{\mathcal{A}(r)\mathcal{B}(r)}{\rm d}r\nonumber\\
&\approx-E_j(p)(T-I_1)-\frac{m_j^2}{2E_j(p)}I_2,
\end{align}
where we have set, for a later convenience,
\begin{align}\label{tI1I2}
    T=\int_A^B{\rm d}t,\qquad I_1=\int_A^B\sqrt{\frac{\mathcal{B}(r)}{\mathcal{A}(r)}}\,{\rm d}r,\qquad I_2=\int_A^B\sqrt{\mathcal{A}(r)\mathcal{B}(r)}\,{\rm d}r.
\end{align}
Note that only when $\mathcal{A}(r)=\mathcal{B}^{-1}(r)$, which is the case in the exterior Schwarzschild solution, does the result of the integral $I_2$ coincide with the coordinate distance $r_B-r_A$. This fact is very important as we will see later. In order to evaluate the Gaussian integral (\ref{CurvedWavePacket}), we need to expand the energy $E_j(p)$ in terms of the momenta $p_j$ as in Eq.\,(\ref{FlatEExpansion}). For that purpose, we use the second identity in Eq.\,(\ref{InfinityObservables}) together with Eq.\,(\ref{massshell}) to find, 
\begin{equation}\label{E(p)inp}
E_j(p)=\sqrt{\frac{\mathcal{A}(r)}{\mathcal{B}(r)}}\left[p_j^2+m_j^2\mathcal{B}(r)\right]^{\frac{1}{2}}.
\end{equation}
This allows us to compute the group velocity $v_i(r)$ and the coefficient $\gamma_i(r)$ in the expansion (\ref{FlatEExpansion}). Note that these are now both position-dependent because of the metric components $\mathcal{A}(r)$ and $\mathcal{B}(r)$ present in Eq.\,(\ref{E(p)inp}). Indeed, using the latter, we find
\begin{equation}\label{vgamma}
    v_j(r)\approx\sqrt{\frac{\mathcal{A}(r)}{\mathcal{B}(r)}}\left(1-\frac{m_j^2\mathcal{B}(r)}{2\bar{p}_j^2}\right),\qquad
    \gamma_j(r)\approx\frac{m_j^2\sqrt{\mathcal{A}(r)\mathcal{B}(r)}}{2\bar{p}_j^3}.
\end{equation}
Inserting the energy expansion (\ref{FlatEExpansion}) into the accumulated phase (\ref{WrongStodolskyPhase}) after setting $r=r_B$ in both $v_j(r)$ and $\gamma_j(r)$ given by (\ref{vgamma}), and plugging the result into the Gaussian integral (\ref{CurvedWavePacket}) by keeping only terms up to the order $\mathcal{O}(m_j^2/\bar{E}_j^3)$, we arrive at the final expression for the wave packet as follows:
\begin{align}\label{WrongWavePacket2}
\psi_j(r_B,r_A;T)&=\int_{-\infty}^{+\infty}\frac{{\rm d}p_j}{2\pi}\left(\frac{2\pi}{\sigma_{pP}^2}\right)^{\frac{1}{4}} \exp\left[-\frac{(p_j-\bar{p}_j)^2}{4\sigma_{pP}^2}-i\left(T-I_1\right)E_j(p)-i\frac{m_j^2}{2E_j(p)}I_2\right]\nonumber\\
&=\!\frac{\left(2\sigma_{pP}^2/\pi\right)^{\frac{1}{4}}}{\sqrt{1+4i\sigma_{pP}^2\gamma_j(T-I_1)}}\exp\!\left[-i(T-I_1)\bar{E}_j-i\frac{m_j^2}{2\bar{E}_j}I_2\!-\!\frac{\sigma_{pP}^2v_j^2\left(\frac{m_j^2}{2\bar{E}_j^2}I_2-(T-I_1)\right)^{\!\!2}}{1+4i\sigma_{pP}^2\gamma_j(T-I_1)}\right]\!.
\end{align}

In order to extract now the flavor transition probability, we need to use this wave packet to compute an amplitude analogous to the amplitude (\ref{ComputedFlatAmplitude}) we wrote in Minkowski spacetime. However, the prescription (\ref{ComputedFlatAmplitude}) is not valid for our wave packet (\ref{WrongWavePacket2}). This is because Eq.\,(\ref{ComputedFlatAmplitude}) involves an integration over all space due to the dependence there of the wave packet on the space coordinate $x$, whereas our wave packet (\ref{WrongWavePacket2}) has already been integrated over space and takes its value at the single location $r=r_B$ of the detection process. Therefore, the flavor transition amplitude should now be found simply by projecting the wave packet (\ref{WrongWavePacket2}) onto the detected neutrino state $\ket{\nu_\beta}$. Here, and henceforth, we will replace, for simplicity, $\sigma_{pP}$ by $\sigma_p$ by ignoring the uncertainty due to the detection process. Thus, according to the wave packet (\ref{WrongWavePacket2}) the squared magnitude of the flavor transition amplitude reads,
\begin{align}\label{WrongCurvedAmplitude2}
    &|\braket{\nu_\beta|\nu_\alpha(r_B,r_A;T)}|^2=\sum_{j,k}\frac{U_{\beta j}U^*_{\alpha j}U_{\beta k}^*U_{\alpha k}\left(2\sigma_{p}^2/\pi\right)^{\frac{1}{2}}}{\sqrt{\left[1+4i\sigma_{p}^2\gamma_j(T-I_1)\right]\left[1-4i\sigma_{p}^2\gamma_k(T-I_1)\right]}}\nonumber\\
    &\times\!\exp\!\left[\!-i(T-I_1)\bar{E}_{jk}\!-\!i\left(\frac{m_j^2}{2\bar{E}_j}-\frac{m_k^2}{2\bar{E}_k}\right)I_2\!-\!\frac{\sigma_{p}^2v_j^2\left(\frac{m_j^2}{2\bar{E}_j^2}I_2\!-\!(T-I_1)\right)^2}{1+4i\sigma_{p}^2\gamma_j(T-I_1)}\!-\!\frac{\sigma_{p}^2v_k^2\left(\frac{m_k^2}{2\bar{E}_k^2}I_2\!-\!(T-I_1)\right)^2}{1-4i\sigma_{p}^2\gamma_k(T-I_1)}\right]\!\!.
\end{align}
Here, we introduced the short-hand notation $\bar{E}_{jk}\equiv\bar{E}_j-\bar{E}_k$. Note that if we limited our energy expansion (\ref{FlatEExpansion}) to the first power of $p_j-\bar{p}_j$, we would have set $\gamma_j=0$ in the result (\ref{WrongCurvedAmplitude2}), in which case we recover the result (59) of Ref.\,\cite{ChatelainVolpe} for the exterior Schwarzschild solution by setting $\mathcal{A}(r)=\mathcal{B}^{-1}(r)=1-2GM/r$. 

As the last step now requires an integration over the unmeasured time $T$, we need to make an approximation similar to the one we made in Minkowski spacetime to get rid of $T$ in the denominators of Eq.\,(\ref{WrongCurvedAmplitude2}). For that purpose, however, we cannot just set $T\approx r_B-r_A$ for ultra-relativistic neutrinos as we did in flat spacetime, but should rather rely on a null geodesic extracted from the general metric (\ref{GeneralMetric}). Using the latter, we easily see that for radially propagating neutrinos we should then set $T\approx I_1$ in the denominators of Eq.\,(\ref{WrongCurvedAmplitude2}). This, as it happens, simply eliminates the imaginary term from those denominators. This is the price to pay for having integrated the Stodolsky phase first before building the wave packet. Indeed, the information about the amplitude decrease and about the widening of the wave packet over time as the latter propagates towards the detector is thus completely lost.

Next, after identifying the energies of the eigenstates with the average energy $\bar{E}_j\approx\bar{E}_k\approx\bar{E}$ and integrating Eq.\,(\ref{WrongCurvedAmplitude2}) over $T$, we arrive at the the following flavor transition probability:
\begin{equation}\label{WrongCurvedProbability}
    \mathcal{P}(\alpha\rightarrow\beta)\propto\sum_{j,k}U_{\beta j}U^*_{\alpha j}U_{\beta k}^*U_{\alpha k}\exp\left(-\frac{2\pi i I_2}{L_{\rm osc}}-\frac{I_2^2}{L_{\rm coh}^2}\right),
\end{equation}
where the oscillation length $L_{\rm osc}$ is as given in Eq.\,(\ref{FlatLengths}), but the coherence length is now given by,
\begin{equation}\label{CurvedLengths2}
    L_{\rm coh}=\frac{4\bar{E}^2\sigma_x}{|\Delta m_{jk}^2|}\frac{\sqrt{v_j^2+v_k^2}}{v_jv_k}.
\end{equation}
We have introduced here, for convenience, the width $\sigma_x=(2\sigma_p)^{-1}$ in coordinate space. The right-hand side in expression (\ref{CurvedLengths2}) is real and thus represents only the damping caused by the mass differences of the mass eigenstates. Using the expression (\ref{vgamma}) for $v_j(r_B)$ and $v_k(r_B)$, as well as the expression (\ref{pinE(p)}) of $p_j$ in terms of $E_j(p)$, we may rewrite this coherence length as,
\begin{equation}\label{ClarifiedCurvedLengths1}
    L_{\rm coh}\approx\frac{4\sqrt{2}\bar{E}^2\sigma_x}{|\Delta m_{jk}^2|}\sqrt{\frac{\mathcal{B}(r_B)}{\mathcal{A}(r_B)}}\left(1+\frac{m_j^2+m_k^2}{4\bar{E}^2}\mathcal{A}(r_B)\right).
\end{equation}

The first thing we notice in this result is that not only the imaginary term that emerges in Minkowski spacetime is absent here, but even the real part does not coincide with the flat-space oscillation length (\ref{FlatLengths}) when we set $\mathcal{A}(r)=\mathcal{B}(r)=1$. Thus, the flavor transition probability (\ref{WrongCurvedProbability}) does not even reduce to the flat-space result (\ref{ComputedFlatProbability}). The second important thing we notice is that what determines the flavor transition probability (\ref{WrongCurvedProbability}) is not the coordinate length $L=r_B-r_A$ as it is the case in the exterior Schwarzschild solution \cite{ChatelainVolpe} for which $\mathcal{A}(r)\mathcal{B}(r)=1$, but rather the integral $I_2$ given by Eq.\,(\ref{tI1I2}). This remark is also valid for the damping term in the exponential (\ref{WrongCurvedProbability}) as it is proportional to $I_2^2$ instead of $L^2$.

\subsection{Building the wave packet first}\label{sec:WavePacketFirst}
As discussed in the Introduction, the main issue not addressed by the previous approach is the effect of curved spacetime on the wave packet itself. In fact, we can now explicitly see that the issue with the previous approach is that by keeping the argument $-\frac{(p_j-\bar{p}_j)^2}{4\sigma_p^2}$ of the Gaussian function outside the phase integral $\Phi_{AB}$ as in Eq.\,(\ref{CurvedWavePacket}) means that one is implicitly assuming that this argument does not depend on the spacetime position. However, when expressed in terms of the conserved quantity $\bar{E}_j(p)$ (using Eq.\,(\ref{pinE(p)})) the mean neutrino momentum $\bar{p}_j$ in the Gaussian function does depend on $\mathcal{A}(r)$ and $\mathcal{B}(r)$, and therefore cannot be factored out of the phase integral as done in Eq.\,(\ref{CurvedWavePacket}). Factoring out the Gaussian function as done in Eq.\,(\ref{CurvedWavePacket}) means, physically, that one allows each constituent plane wave of the wave packet to ``propagate"\footnote{Of course, the plane waves making a wave packet are, more properly, viewed as non-propagating. What propagates is the wave packet that emerges from such a superposition of non-propagating plane waves.} without interfering with each other along the neutrino's path from the source to the detector. In other words, only when the detector is reached by all the plane waves that the latter are combined to build a self-interfering wave packet. In so doing, the decrease in amplitude and the widening caused to the wave packet by gravity along its journey would be completely missed  as we just saw in the previous section. 

The alternative approach consists then in evaluating first the Gaussian integral (\ref{FlatWavePacket}) to build a wave packet, and then evaluate the effect of spacetime on the wave packet by finding the form of the latter at the detection point. Inserting the expansion (\ref{FlatEExpansion}) with the position-dependent $v_i(r)$ and $\gamma_i(r)$ into the Gaussian integral (\ref{FlatWavePacket}), we find the following expression for the wave packet:
\begin{equation}\label{FinalCurvedWP}
    \psi_j(r,t)=\frac{(2\sigma_p^2/\pi)^{\frac{1}{4}}}{\sqrt{1+4i\sigma_p^2\gamma_j(r)t}}\exp\left[-i\bar{E}_jt+i\bar{p}_jr-\frac{\sigma_p^2(r-v_j(r)t)^2}{1+4i\sigma_p^2\gamma_j(r)t}\right].
\end{equation}
This expression gives us a form for the wave packet that is valid at any coordinate $r$ of space and for all times $t$. We only need then to find its expression at the detection point and then project it onto the detected neutrino state $\ket{\nu_
\beta}$\footnote{Note that the reason why we discard here the wave packet nature of the detected neutrino is that we work now with the radial coordinate $r$ instead of $x$. As such, an integration over $r$ (as in Eq.\,(\ref{ComputedFlatAmplitude})) would not lead to a simple Gaussian integral as in Minkowski spacetime. In addition, to make our integral over $r$ covariant, we would need to introduce $\sqrt{g_{rr}}$ into the integration measure. This would not lead to any easy way for computing the $r$-integral. Thanks to the Gaussian function, however, which peaks at the detection point, we may safely just take the detected neutrino to be the Fock state $\ket{\nu_\beta}$ as we did in the previous section without losing the effect of the curved spacetime.}. However, what we have now is a {\it propagating} entity. As such, it might still seem illegitimate to compute the accumulated phase by integrating independently over $r$ and $t$ and discard the link (\ref{dr/dt}) between these two coordinates.
The reason why this procedure is consistent now is that by building the wave packet first, one fully preserves the meaning of the unmeasured time $T$ which has nothing to do now with a propagation of on-shell mass eigenstates. 

\begin{figure}[H]
    \centering
    \includegraphics[scale=0.66]{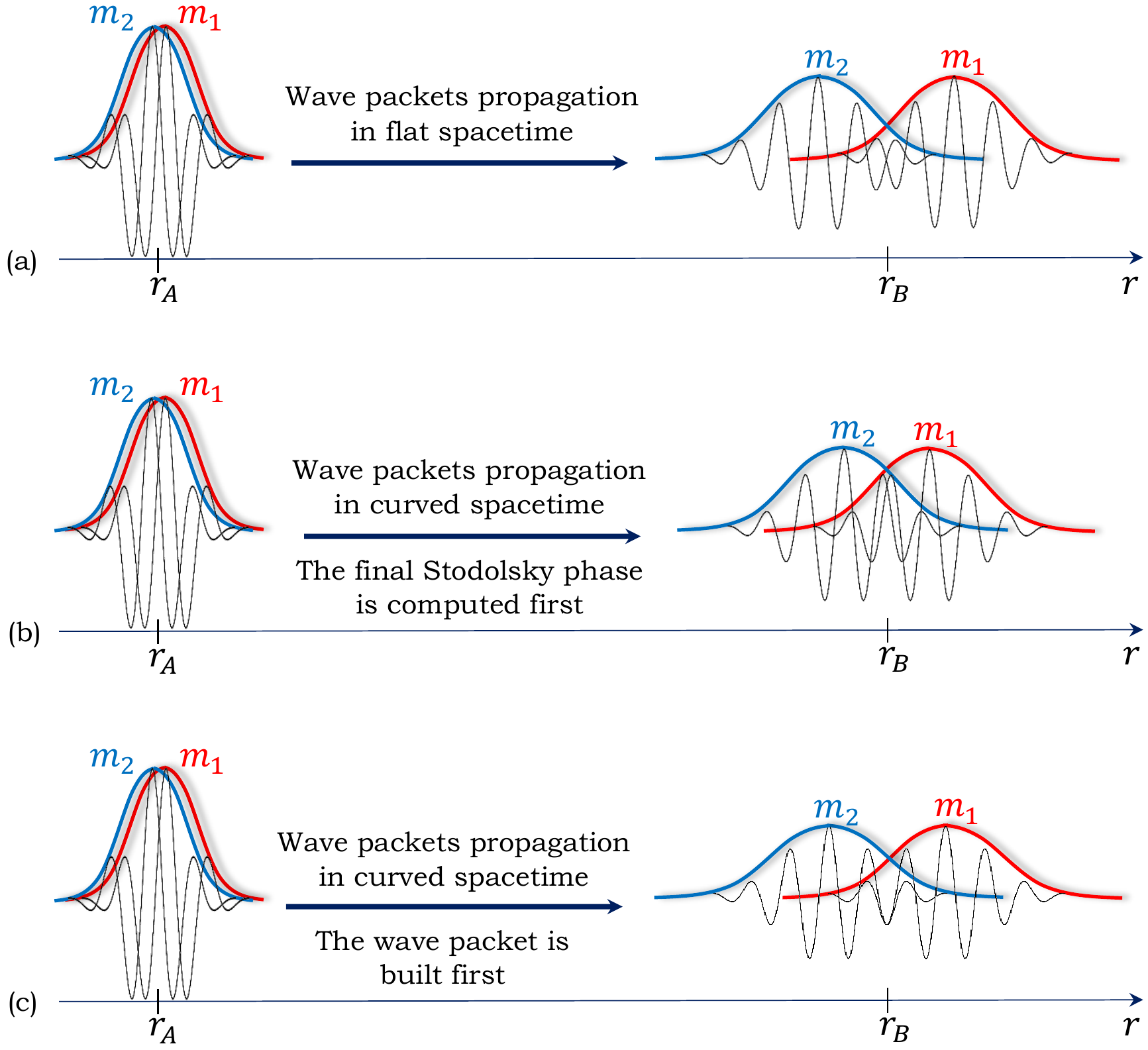}
    \caption{\small The spreading of two neutrino-mass-eigenstates wave packets are shown in (a) flat spacetime and in [(b),(c)] curved spacetime. The coherence length (thus, the amount of overlap) in (b), where the wave packet is built after computing the final Stodolsky phase, is larger (see Eq.\,(\ref{ClarifiedCurvedLengths1})) than in (c), where the wave packet is built first before evaluating the final phase (see Eq.\,(\ref{CoherenceLength})). Similarly, the coherence length in (c) is larger than in (a) (compare Eqs.\,(\ref{ZSplit}) and (\ref{CurvedZSplit})). The relative final phase of the oscillating waves under the envelops in (a) is also different from the one in (c) (compare Eqs.\,(\ref{ClarifiedFlatProbability}) and (\ref{FinalCorrectCurvedProbability})).}
    \label{Figure}
\end{figure}

We thus first compute the accumulated quantum phase of the wave packet (\ref{FinalCurvedWP}) \`a la Stodolsky by still integrating independently over $t$ and $r$ to get exactly the result Eq.\,(\ref{WrongStodolskyPhase}), but with $\bar{E}_j(p)$ there replaced by $\bar{E}_j$. For the rest, we will set $r=L$ in the terms which do not contribute to the phase and set, again as an approximation, $t\approx I_1$ in the complex denominators. In the numerator of the last term inside the exponential, however, we only set $t=T$ to allow it to be integrated over to extract the probability. With these replacements, the wave packet (\ref{FinalCurvedWP}) takes the following form at the detection point:
\begin{equation}\label{CorrectWavePacket2}
    \psi_j(r_B,r_A;T)=\frac{(2\sigma_p^2/\pi)^{\frac{1}{4}}}{\sqrt{1+4i\sigma_p^2\gamma_jI_1}}\exp\left[-i\bar{E}_j(T-I_1)-i\frac{m_j^2}{2\bar{E}_j}I_2-\frac{\sigma_p^2(L-v_jT)^2}{1+4i\sigma_p^2\gamma_jI_1}\right].
\end{equation}
Therefore, the squared magnitude of the flavor transition amplitude reads,
\begin{multline}\label{CorrectCurvedAmplitude2}
    |\braket{\nu_\beta|\nu_\alpha(r_B,r_A;T)}|^2=\sum_{j,k}\frac{U_{\beta j}U^*_{\alpha j}U_{\beta k}^*U_{\alpha k}(2\sigma_p^2/\pi)^{\frac{1}{2}}}{\sqrt{\left(1+4i\sigma_{p}^2\gamma_jI_1\right)\left(1-4i\sigma_{p}^2\gamma_kI_1\right)}}\\
    \times\exp\left[-i(T-I_1)\bar{E}_{jk}-i\left(\frac{m_j^2}{2\bar{E}_j}-\frac{m_k^2}{2\bar{E}_k}\right)I_2-\frac{\sigma_p^2(L-v_jT)^2}{1+4i\sigma_p^2\gamma_jI_1}-\frac{\sigma_p^2(L-v_kT)^2}{1-4i\sigma_p^2\gamma_kI_1}\right].
\end{multline}
Note that, in contrast to what happens to the result (\ref{WrongCurvedAmplitude2}) when setting $T\approx I_1$ in the denominators, the imaginary terms in the latter still survive here. This is the advantage one gains by building the wave packet right from the neutrino's source before working out its accumulated quantum phase. All the information about the amplitude decrease and about the spreading of the wave packet over time as it propagates towards the detector is completely preserved (see Fig.\,\ref{Figure}).

Next, after identifying the energies of the eigenstates with the common average energy $\bar{E}_j\approx\bar{E}_k\approx\bar{E}$ and integrating Eq.\,(\ref{CorrectCurvedAmplitude2}) over $T$, we arrive at the following flavor transition probability:
\begin{equation}\label{CorrectCurvedProbability2}
    \mathcal{P}(\alpha\rightarrow\beta)\propto\sum_{j,k}U_{\beta j}U^*_{\alpha j}U_{\beta k}^*U_{\alpha k}\exp\left[-2\pi i\left(\frac{I_2}{L_{\rm osc}}+\frac{1}{4\pi}\tan^{-1}\left[\frac{(\gamma_j-\gamma_k)I_1}{\sigma_x^2}\right]\right)-\frac{L^2}{\mathcal{Z}_{\rm coh}^2}\right].
\end{equation}
Here, $L_{\rm osc}$ and $\mathcal{Z}_{\rm coh}$ have exactly the same expressions as their flat-space counterparts (\ref{FlatLengths}). However, the content inside the term $\mathcal{Z}_{\rm coh}$ is here slightly different in that the complex width $\tilde{\sigma}_p$ in expression (\ref{TildeSigma}) comes now with a different imaginary part:
\begin{equation}\label{CurvedTildeSigma}
    \frac{1}{\tilde{\sigma}_{pj}^2}=\frac{1}{\sigma_{p}^2}+4i\gamma_j(r_B)I_1.
\end{equation}
Using the expressions of $v_j(r_B)$ and $\gamma_j(r_B)$ as given by Eq.\,(\ref{vgamma}), as well as the expression (\ref{pinE(p)}) of $p_j$ in terms of $E_j(p)$, we find up to the leading order in $\Delta m_{jk}^2/\bar{E}^2$ the following:
\begin{equation}\label{CurvedZSplit}
\mathcal{Z}_{\rm coh}=\frac{4\sqrt{2}\bar{E}^2\sigma_x}{|\Delta m_{jk}^2|}\mathcal{A}^{-1}(r_B)\left(1-\frac{m_j^2+m_k^2}{4\bar{E}^2}\mathcal{A}(r_B)\right)+\frac{iI_1}{\sigma_x\bar{E}\sqrt{2}}\frac{\mathcal{A}(r_B)}{\mathcal{B}(r_B)}.
\end{equation}
In contrast to the result (\ref{ClarifiedCurvedLengths1}), this expression contains an imaginary part and reduces exactly to the flat-space result (\ref{ZSplit}) when $\mathcal{A}(r)=\mathcal{B}(r)=1$. We also recognize a neat difference between the real part in this expression and the real part of the result (\ref{ClarifiedCurvedLengths1}) found using the first approach. The real part, which represents the coherence length $L_{\rm coh}$, is here smaller as it consists of a difference:
\begin{equation}\label{CoherenceLength}
    L_{\rm coh}\approx\frac{4\sqrt{2}\bar{E}^2\sigma_x}{|\Delta m_{jk}^2|}\mathcal{A}^{-1}(r_B)\left(1-\frac{m_j^2+m_k^2}{4\bar{E}^2}\mathcal{A}(r_B)\right).
\end{equation}
In addition, in contrast to the result (\ref{WrongCurvedProbability}), expression (\ref{CorrectCurvedProbability2}) involves both integrals $I_1$ and $I_2$ and reduces exactly to the flat-space transition probability (\ref{ComputedFlatProbability}). This is a remarkable result which is very relevant to modified gravity theories. In fact, this is what will allow the emergence of a rich distinction between various spacetime metrics when we apply our results using explicit expressions of $\mathcal{A}(r)$ and $\mathcal{B}(r)$, starting from Sec.\,\ref{sec:ExteriorSchw}. We may also express the coherence length (\ref{CoherenceLength}) in terms of the locally measured average energy $\bar{E}_{\rm loc}=[\mathcal{A}(r_B)]^{-\frac{1}{2}}\bar{E}$, as follows:
\begin{equation}\label{LocalCoherenceLength}
L_{\rm coh}\approx\frac{4\sqrt{2}\bar{E}_{\rm loc}^2\sigma_x}{|\Delta m_{jk}^2|}\left(1-\frac{m_j^2+m_k^2}{4\bar{E}_{\rm loc}^2}\right).
\end{equation}

To see now explicitly the effect of combining gravity with the wave-packet nature of the neutrinos on their flavor transition probability, we plug expression (\ref{CurvedZSplit}) as well as Eq.\,(\ref{vgamma}) into the exponential in Eq.\,(\ref{CorrectCurvedProbability2}) so that the latter takes the form,
\begin{multline}\label{FinalCorrectCurvedProbability}
\mathcal{P}(\alpha\rightarrow\beta)\propto\sum_{j,k}U^*_{\alpha j}U_{\beta j}U_{\alpha k}U^*_{\beta k}\\
\times\exp\left[-2\pi i\left(\frac{I_2}{L_{\rm osc}}+\frac{1}{4\pi}\tan^{-1}\left[\frac{2\pi}{\sigma_x^2\bar{E}^2_{\rm loc}}\frac{\mathcal{A}(r_B)I_1}{\mathcal{B}(r_B)L_{\rm osc}}\right]-\frac{\sqrt{2}I_1}{2\pi\sigma_x\bar{E}_{\rm loc}}\frac{\sqrt{\mathcal{A}(r_B)}L^2}{\mathcal{B}(r_B)L^3_{\rm coh}}\right)-\frac{L^2}{L_{\rm coh}^2}\right].
\end{multline}
When we compare this expression to the flat-space result (\ref{ClarifiedFlatProbability}), we see that both additional terms that contribute to the flavor transition phase in flat space are here corrected by factors that depend on the separate metric components $\mathcal{A}(r_B)$ and $\mathcal{B}(r_B)$ as well as on the integrals $I_1$ and $I_2$. Of course, the coordinate distance $L=r_B-r_A$ displayed in Eq.\,(\ref{FinalCorrectCurvedProbability}) may be expressed in terms of the proper distance $L_p=\int_A^B\sqrt{\mathcal{B}(r)}{\rm d}r$ after fixing the specific form of the metric components. 

In the next sections we will apply the result (\ref{FinalCorrectCurvedProbability}) to various metrics by computing the corresponding integrals $I_1$ and $I_2$ for each case. We are not going to apply the result (\ref{WrongCurvedProbability}) obtained within the first approach since the latter is unable to take into account the widening of the wave packet and since, as we saw, it is not even able to reproduce the flat-space result (\ref{ComputedFlatProbability}).
\section{In the exterior Schwarzschild solution}\label{sec:ExteriorSchw}
In this section, we apply the result we obtained in the previous section to the case of the Schwarzschild spacetime outside a massive spherical object of mass $M$. In this case, the metric components $\mathcal{A}(r)$ and $\mathcal{B}(r)$ are given, up to the first order in $GM$ by \cite{Synge},
\begin{equation}\label{InSchwarzschild}
    \mathcal{A}(r)=1-\frac{2GM}{r},\qquad \mathcal{B}(r)\approx1+\frac{2GM}{r}.
\end{equation}
Using these functions of $r$, we compute, up to the first order in $GM$, the integrals $I_1$ and $I_2$ given by (\ref{tI1I2}) as follows,
\begin{align}\label{I1I2ForIntSchw}
I_1&\approx L+2GM\ln\left(\frac{r_B}{r_A}\right)\approx L_p+GM\ln\left(\frac{r_B}{r_A}\right),\nonumber\\
I_2&= L\approx L_p-GM\ln\left(\frac{r_B}{r_A}\right).
\end{align}
In the second steps we have expressed the coordinate distance $L$ in terms of the proper distance $L_p$. Plugging these expressions into Eq.\,(\ref{FinalCorrectCurvedProbability}), we find, up to the first order in $GM$, the following oscillation phase:
\begin{multline}\label{ExtSchwPhase}
\varphi_{jk}=\frac{2\pi L_p}{L_{\rm osc}}\left[1-\frac{GM}{L_p}\ln\left(\frac{r_B}{r_A}\right)+\frac{GM}{r_B}\right]+\frac{1}{2}\tan^{-1}\left(\frac{2\pi L_p}{\sigma_x^2\bar{E}^2_{\rm loc}L_{\rm osc}}\left[1+\frac{GM}{L_p}\ln\left(\frac{r_B}{r_A}\right)-\frac{3GM}{r_B}\right]\right)\\
-\frac{\sqrt{2}}{\sigma_x\bar{E}_{\rm loc}}\frac{L_p^3}{L^3_{\rm coh}}\left[1-\frac{GM}{L_p}\ln\left(\frac{r_B}{r_A}\right)-\frac{3GM}{r_B}\right].
\end{multline}
Here (and henceforth), both $L_{\rm osc}$ and $L_{\rm coh}$ are expressed in terms of the locally measured average energy $\bar{E}_{\rm loc}$. We clearly see from this expression the corrections brought by the curved space to the oscillation phase found in flat space. The first square brackets represent the correction brought to the plane-wave phase. The inverse tangent represents the correction brought to the transition probability, which is an enhancement effect due to the widening of the wave packets of each mass eigenstate. The last square brackets represent the effect of the wave packets dispersion. On the other hand, the damping term in the exponential in Eq.\,(\ref{FinalCorrectCurvedProbability}) is given in terms of the proper length $L_p$ as follows,
\begin{equation}\label{ExtSchwDamp}
\exp\left(-\frac{L^2}{L^2_{\rm coh}}\right)    \approx\exp\left(-\frac{L_p^2}{L^2_{\rm coh}}\left[1-\frac{2GM}{L_p}\ln\left(\frac{r_B}{r_A}\right)\right]\right).    
\end{equation}
From Eqs.\,(\ref{ExtSchwPhase}) and (\ref{ExtSchwDamp}) we see that gravity reduces the damping caused by the loss of coherence when $r_B>r_A$ ({\it i.e.}, when the neutrinos climb the gravitational potential) and increases the damping when $r_B<r_A$ ({\it i.e.}, when the neutrinos delve into the gravitational potential).
\section{In the interior Schwarzschild solution}\label{sec:InteriorSchw}
Since neutrinos travel inside the Earth (and also inside other astrophysical objects) before they reach the detector, it is very important to examine here the effect of the gravitational field inside massive objects which we take here, for simplicity, to be spherically symmetric and of uniform mass density. The metric components $\mathcal{A}(r)$ and $\mathcal{B}(r)$ of the interior Schwarzschild solution of such an object of mass $M$ and of radius $R$ are given by \cite{Synge},
\begin{equation}\label{IntSchwarzschild}
    \mathcal{A}(r)\approx1-\frac{3GM}{R}+\frac{GMr^2}{R^3},\qquad \mathcal{B}(r)\approx1+\frac{2GMr^2}{R^3}.
\end{equation}
Therefore, we compute, up to the first order in $GM$, the integrals $I_1$ and $I_2$ using Eq.\,(\ref{tI1I2}) as follows,
\begin{align}\label{I1I2ForExteriorSchw}
I_1&\approx D+\frac{10GM}{3}\approx D_p+\frac{8GM}{3},\nonumber\\
I_2&\approx D-2GM \approx D_p-\frac{8GM}{3}.    
\end{align} 
We have assumed here that the neutrino flux traverses the sphere diametrically by going through the center of the sphere. We have thus performed the integration by taking $r_A=0$ and $r_B=R$ and then multiplied the result by a factor of two. We denoted by $D$ the coordinate diameter of the sphere and by $D_p$ the proper length of the diameter.

Plugging these expressions into Eq.\,(\ref{FinalCorrectCurvedProbability}), we find, up to the first order in $GM$, the following oscillation phase:
\begin{equation}\label{IntSchwPhase}
\varphi_{jk}=\frac{2\pi D_p}{L_{\rm osc}}\left(1-\frac{2GM}{3D_p}\right)+\frac{1}{2}\tan^{-1}\left(\frac{2\pi D_p}{\sigma_x^2\bar{E}^2_{\rm loc}L_{\rm osc}}\left[1-\frac{10GM}{3D_p}\right]\right)-\frac{\sqrt{2}}{\sigma_x\bar{E}_{\rm loc}}\frac{D_p^3}{L^3_{\rm coh}}\left(1-\frac{14GM}{3D_p}\right).\end{equation}
As there are no logarithms in this expression, we have a simpler correction than the one caused by the exterior Schwarzschild solution. Similarly, the damping term in the exponential in Eq.\,(\ref{FinalCorrectCurvedProbability}) is given in terms of the proper length $D_p$ of the diameter as follows,
\begin{equation}\label{IntSchwDamp}
\exp\left(-\frac{L^2}{L^2_{\rm coh}}\right)    \approx\exp\left(-\frac{D_p^2}{L^2_{\rm coh}}\left[1-\frac{4GM}{3D_p}\right]\right).    
\end{equation} 
This formulas gives the effect of pure gravity without taking into account the interaction of the neutrinos with the matter background.

Nevertheless, we may easily include the effect of matter in this result as done in Ref.\,\cite{Buoninfante} by following Ref.\,\cite{Cardall} for taking into account the effect on the flavor oscillations caused by interactions with matter. Note, however, that unlike the prescription to deal with matter effects given in Eqs.\,(65) and (66) of Ref.\,\cite{Buoninfante}, our prescription here for our wave packet treatment consists simply in the replacement $\Delta m^2\rightarrow\Delta\mu^2$, where the effective squared mass $\mu^2$ is obtained by diagonalizing the effective mass matrix $M^2_f-V_f$. In fact, for a 2-flavor neutrino oscillation (for simplicity), we have in the flavor basis the following mixing matrix, vacuum mass matrix and effective four-vector potential matrix for the interaction with the electron background, respectively \cite{Cardall}:
\begin{equation}
    U=\begin{pmatrix}
\cos\vartheta & \sin\vartheta\\
-\sin\vartheta & \cos\vartheta
\end{pmatrix},\qquad M_f^2=U\begin{pmatrix}
m_1^2 & 0\\
0 & m_2^2
\end{pmatrix}U^\dagger,\qquad A^\mu_f=\begin{pmatrix}
-\sqrt{2}G_Fn_e u^\mu & 0\\
0 & 0
\end{pmatrix}. 
\end{equation} 
Here, $G_F$ is the Fermi constant, $n_e$ is the rest-frame density of electrons and $u^\mu$ is the four-velocity of the electron fluid. The mass-shell relation for the neutrinos inside matter may then be written as \cite{Cardall}: $-M_f^2=g_{\mu\nu}\left(p_j^\mu+A_f^\mu\mathcal{P}_L\right)\left(p_j^\nu+A_f^\nu\mathcal{P}_L\right)$, where $\mathcal{P}_L$ is the left-handed projection operator. Therefore, assuming that the electron background is at rest with respect to the oscillation experiment and keeping only terms to first order in $G_F$, this mass-shell condition leads to the following modification of the vacuum relation (\ref{pinE(p)}):
\begin{equation}\label{pinE(p)+MatterBackground}
p=E(p)\sqrt{\frac{\mathcal{B}(r)}{\mathcal{A}(r)}}\left[1-\frac{(M_f^2-V_f)\mathcal{A}(r)}{E^2(p)}\right]^{\frac{1}{2}},
\end{equation}
where the effective mass contribution from the matter background is given by,
\begin{equation}\label{Vf}
    V_f=\begin{pmatrix}
-2\sqrt{2}G_FE(p)n_e\mathcal{P}_L & 0\\
0 & 0
\end{pmatrix}. 
\end{equation} 
Thus, to include the effect of matter in our results (\ref{IntSchwPhase}) and (\ref{IntSchwDamp}), we only need to extract the effective masses to use for the mass eigenstates after diagonalizing the effective mass matrix $M_f^2-V_f$. 
\section{In de Sitter-Schwarzschild spacetime}\label{sec:deSitterSchw}
In this section we apply our results to the case of a gravitational field created by spherically symmetric massive objects inside an expanding universe. For that purpose, we use the metric components of the de Sitter-Schwarzschild spacetime. The plane wave treatment of neutrino oscillations in such a spacetime has been conducted in Refs.\,\cite{JunPen,Tao}. The metric components $\mathcal{A}(r)$ and $\mathcal{B}(r)$ in static coordinates are given by \cite{Synge},
\begin{equation}\label{deSitterSchw}
    \mathcal{A}(r)=1-\frac{2GM}{r}-H^2 r^2,\qquad \mathcal{B}(r)\approx1+\frac{2GM}{r}+H^2 r^2.
\end{equation}
Here, $H$ stands for a constant Hubble parameter that is related to the cosmological constant $\Lambda$ by $H=\sqrt{\Lambda/3}$\footnote{These metric components are also very useful for studying the effect of the cosmic expansion on the bending of light by massive objects. See, e.g., Ref.\,\cite{LightBending} and the references therein.}. Using these components, we compute, up to the first order in $GM$ and $H^2$, the integrals $I_1$ and $I_2$ using Eq.\,(\ref{tI1I2}) as follows,
\begin{align}\label{I1I2deSitterSchw}
I_1&\approx L+2GM\ln\left(\frac{r_B}{r_A}\right)+\frac{H^2}{3}\left(r_B^3-r_A^3\right)\approx L_p+GM\ln\left(\frac{r_B}{r_A}\right)+\frac{H^2}{6}\left(r_B^3-r_A^3\right),\nonumber\\
I_2&=L\approx L_p-GM\ln\left(\frac{r_B}{r_A}\right)-\frac{H^2}{6}\left(r^3_B-r^3_A\right).    
\end{align} 
Plugging these expressions into Eq.\,(\ref{FinalCorrectCurvedProbability}), we find, up to the first order in $GM$ and $H^2$, the following oscillation phase:
\begin{multline}\label{deSitterSchwPhase}
\varphi_{jk}=\frac{2\pi L_p}{L_{\rm osc}}\left[1-\frac{GM}{L_p}\ln\left(\frac{r_B}{r_A}\right)-\frac{H^2}{6L_p}\left(r_B^3-r_A^3\right)+\frac{GM}{r_B}+\frac{H^2r_B^2}{2}\right]\\
+\frac{1}{2}\tan^{-1}\left(\frac{2\pi L_p}{\sigma_x^2\bar{E}^2_{\rm loc}L_{\rm osc}}\left[1+\frac{GM}{L_p}\ln\left(\frac{r_B}{r_A}\right)+\frac{H^2}{6L_p}\left(r_B^3-r_A^3\right)-\frac{3GM}{r_B}-\frac{3H^2r_B^2}{2}\right]\right)\\
-\frac{\sqrt{2}}{\sigma_x\bar{E}_{\rm loc}}\frac{L_p^3}{L^3_{\rm coh}}\left[1-\frac{GM}{L_p}\ln\left(\frac{r_B}{r_A}\right)-\frac{H^2}{6L_p}\left(r_B^3-r_A^3\right)-\frac{3GM}{r_B}-\frac{3H^2r_B^2}{2}\right].
\end{multline}
The damping term in the exponential in Eq.\,(\ref{FinalCorrectCurvedProbability}) is given in terms of the proper length $L_p$ as follows,
\begin{equation}\label{deSitterSchwDamp}
\exp\left(-\frac{L^2}{L^2_{\rm coh}}\right)    \approx\exp\left(-\frac{L_p^2}{L^2_{\rm coh}}\left[1-\frac{2GM}{L_p}\ln\left(\frac{r_B}{r_A}\right)-\frac{H^2}{3L_p}\left(r_B^3-r_A^3\right)\right]\right).    
\end{equation} 
From these results, we clearly see that the effect of the cosmic expansion is to decrease the oscillation phase as well as to decrease the damping effect.
\section{Application to selected modified gravity models}\label{sec:ModifiedGravity}
In this section, we apply our results to different metrics obtained for specific modified gravity models from the literature. The selected metrics that we are going to examine here are the ones already studied in Ref.\,\cite{Buoninfante} based on the plane wave treatment of neutrino oscillations. The goal of this section is therefore to find out what difference the wave-packet nature of neutrinos could bring to distinguish between the various modifications to the Newtonian potential displayed by the metric of a curved spacetime.   
\subsection{$f(\mathcal{R})$-gravity}\label{sec:ModelI}
The first model we pick up belongs to the so-call $f(\mathcal{R})$-gravity models class. The gravitational Lagrangian in these models is an arbitrary functional of the Ricci scalar $\mathcal{R}$. However, we will only focus here, as done in Ref.\,\cite{Buoninfante}, on the simple quadratic model $f(\mathcal{R})=\mathcal{R}+\alpha\mathcal{R}^2$ \cite{Sakharov}, where the free parameter $\alpha$ is constrained from observations. This model is of great importance for cosmology as it is not only consistent with observations, but it is also able to describe inflation without requiring an inflaton field \cite{Starobinsky}. The metric components that emerge from this model are of the form:
\begin{equation}\label{f(R)}
    \mathcal{A}(r)\approx1-\frac{2GM}{r}\left(1+\frac{e^{-m_0r}}{3}\right),\qquad \mathcal{B}(r)\approx1+\frac{2GM}{r}\left(1-\frac{e^{-m_0r}}{3}\right).
\end{equation}
The constant $m_0$ is related to the constant $\alpha$ by $m_0=\sqrt{2/3\alpha}$.
Using these components of the metric, we compute, up to the first order in $GM$, the integrals $I_1$ and $I_2$ given by  Eq.\,(\ref{tI1I2}) as follows,
\begin{align}\label{I1I2Model1}
I_1&\approx  L+2GM\ln\left(\frac{r_B}{r_A}\right)\approx L_p+GM\ln\left(\frac{r_B}{r_A}\right)+\frac{GM}{3}\left[{\rm Ei}(-m_0r)\right]_{r_A}^{r_B},\nonumber\\
I_2&\approx L-\frac{2GM}{3}\left[{\rm Ei}(-m_0r)\right]_{r_A}^{r_B}\approx L_p-GM\ln\left(\frac{r_B}{r_A}\right)-\frac{GM}{3}\left[{\rm Ei}(-m_0r)\right]_{r_A}^{r_B}.
\end{align}
Here, the function ${\rm Ei}(z)$ is the exponential integral defined by ${\rm Ei}(z)=-\int_{-z}^\infty t^{-1}e^{-t}{\rm d}t$ for any real nonzero variable $z$ \cite{MathHandBook}. Plugging these expressions into Eq.\,(\ref{FinalCorrectCurvedProbability}), we find, up to the first order in $GM$, the following oscillation phase:
\begin{multline}\label{ModelIPhase}
\varphi_{jk}=\frac{2\pi L_p}{L_{\rm osc}}\left[1-\frac{GM}{L_p}\ln\left(\frac{r_B}{r_A}\right)-\frac{GM}{3L_p}\left[{\rm Ei}(-m_0r)\right]_{r_A}^{r_B}+\frac{GM}{r_B}\left(1+\frac{e^{-m_0r_B}}{3}\right)\right]\\
+\frac{1}{2}\tan^{-1}\left(\frac{2\pi L_p}{\sigma_x^2\bar{E}^2_{\rm loc}L_{\rm osc}}\left[1+\frac{GM}{L_p}\ln\left(\frac{r_B}{r_A}\right)+\frac{GM}{3L_p}\left[{\rm Ei}(-m_0r)\right]_{r_A}^{r_B}-\frac{3GM}{r_B}\left(1-\frac{e^{-m_0r_B}}{9}\right)\right]\right)\\
-\frac{\sqrt{2}}{\sigma_x\bar{E}_{\rm loc}}\frac{L_p^3}{L^3_{\rm coh}}\left[1-\frac{GM}{L_p}\ln\left(\frac{r_B}{r_A}\right)+\frac{GM}{L_p}\left[{\rm Ei}(-m_0r)\right]_{r_A}^{r_B}-\frac{3GM}{r_B}\left(1-\frac{e^{-m_0r_B}}{9}\right)\right].
\end{multline}
On the other hand, the damping term in the exponential in Eq.\,(\ref{FinalCorrectCurvedProbability}) is given in terms of the proper length $L_p$ as follows,
\begin{equation}\label{ModelIDamp}
\exp\left(-\frac{L^2}{L^2_{\rm coh}}\right)    \approx\exp\left(-\frac{L_p^2}{L^2_{\rm coh}}\left[1-\frac{2GM}{L_p}\ln\left(\frac{r_B}{r_A}\right)+\frac{2GM}{3L_p}\left[{\rm Ei}(-m_0r)\right]_{r_A}^{r_B}\right]\right).
\end{equation} 
We see from these expressions that the gravitational corrections brought by the modified metric components increase the damping caused by the wave-packet nature of the mass eigenstates. 

The next models discussed in Ref.\,\cite{Buoninfante} are the forth-order \cite{Stelle} and the sixth-order \cite{6Order} gravity models as well as a few nonlocal gravity models. The metric components that emerge from the fourth-order gravity model differ from the components (\ref{f(R)}) only by extra exponential terms of the form $e^{-m_2r}$, where $m_2$ is another constant parameter. The metric components that emerge from the sixth-order gravity model differ from the components that emerge from the fourth-order gravity model by the fact that each exponential comes multiplied by a cosine function, in the manner $e^{-m_0r}\cos(m_0r)$. For this reason, we are not going to examine explicitly those higher-order derivative models here as the results will be very similar to expressions (\ref{ModelIPhase}) and (\ref{ModelIDamp}), only displaying extra terms of the form $GM{\rm Ei}(z)$. For similar reasons, among the non-local gravity models treated in Ref.\,\cite{Buoninfante} we are going to examine in what follows only the infinite-derivative model as it leads to noticeably simple and yet very distinct expressions.

\subsection{Infinite-derivative model}\label{sec:ModelVI}
The second modified gravity model we pick up here is a ghost-free nonlocal gravity model whose quadratic Lagrangian contains infinite-derivative operators. Up to the second order in the metric perturbation \cite{Buoninfante}, the explicit form of the Lagrangian is taken to be $\mathcal{R}+\mathcal{R}f(\Box)\mathcal{R}+\mathcal{R}_{\mu\nu}g(\Box)\mathcal{R}^{\mu\nu}$, where $\mathcal{R}_{\mu\nu}$ is the Ricci tensor and the functions $f$ and $g$ of the d'Alembertian operator $\Box=g^{\mu\nu}\nabla_\mu\nabla_\nu$ are $f(\Box)=-\frac{1}{2}g(\Box)=\frac{e^{-\Box/M_*}-1}{\Box}$ for some energy scale $M_*$ \cite{Infinite,Buoninfante}. The metric components that emerge from such a Lagrangian are of the form \cite{Infinite,Buoninfante},
\begin{equation}\label{InfiniteDerivative}
    \mathcal{A}(r)\approx1-\frac{2GM}{r}{\rm erf}\left(\frac{M_*r}{2}\right),\qquad \mathcal{B}(r)\approx1+\frac{2GM}{r}{\rm erf}\left(\frac{M_*r}{2}\right),
\end{equation}
where, ${\rm erf}(z)=\frac{2}{\sqrt{\pi}}\int_0^ze^{-t^2}{\rm d}t$ is the error function for any real variable $z$ \cite{MathHandBook}. We compute the integrals $I_1$ and $I_2$  using Eq.\,(\ref{tI1I2}) directly in terms of the proper length $L_p$, and up to the first order in $GM$, as follows,
\begin{align}\label{I1I2Infinite}
I_1&\approx L_p+GM\left[\frac{M_*r}{\sqrt{\pi}}\,_2F_2\left(\tfrac{1}{2},\tfrac{1}{2};\tfrac{3}{2},\tfrac{3}{2};-\frac{M_*^2r^2}{4}\right)\right]_{r_A}^{r_B},\nonumber\\
I_2&\approx L_p-GM\left[\frac{M_*r}{\sqrt{\pi}}\,_2F_2\left(\tfrac{1}{2},\tfrac{1}{2};\tfrac{3}{2},\tfrac{3}{2};-\frac{M_*^2r^2}{4}\right)\right]_{r_A}^{r_B}.
\end{align}
Here, $\,_2F_2\left(a,b;c,d;z\right)$ is the generalized hypergeometric functions for any real variable $z$ \cite{MathHandBook}.  Plugging these expressions into Eq.\,(\ref{FinalCorrectCurvedProbability}), we find, up to the first order in $GM$, the following oscillation phase:
\begin{multline}\label{InfinitePhase}
\varphi_{jk}=\frac{2\pi L_p}{L_{\rm osc}}\left\{1-\frac{GM}{L_p}\left[\frac{M_*r}{\sqrt{\pi}}\,_2F_2\left(\tfrac{1}{2},\tfrac{1}{2};\tfrac{3}{2},\tfrac{3}{2};-\frac{M_*^2r^2}{4}\right)\right]_{r_A}^{r_B}+\frac{GM}{r_B}{\rm erf}\left(\frac{M_*r_B}{2}\right)\right\}\\
+\frac{1}{2}\tan^{-1}\left(\frac{2\pi L_p}{\sigma_x^2\bar{E}^2_{\rm loc}L_{\rm osc}}\left[1+\frac{GM}{L_p}\left[\frac{M_*r}{\sqrt{\pi}}\,_2F_2\left(\tfrac{1}{2},\tfrac{1}{2};\tfrac{3}{2},\tfrac{3}{2};-\frac{M_*^2r^2}{4}\right)\right]_{r_A}^{r_B}-\frac{3GM}{r_B}{\rm erf}\left(\frac{M_*r_B}{2}\right)\right]\right)\\
-\frac{\sqrt{2}}{\sigma_x\bar{E}_{\rm loc}}\frac{L_p^3}{L^3_{\rm coh}}\left\{1-\frac{GM}{L_p}\left[\frac{M_*r}{\sqrt{\pi}}\,_2F_2\left(\tfrac{1}{2},\tfrac{1}{2};\tfrac{3}{2},\tfrac{3}{2};-\frac{M_*^2r^2}{4}\right)\right]_{r_A}^{r_B}-\frac{3GM}{r_B}{\rm erf}\left(\frac{M_*r_B}{2}\right)\right\}.
\end{multline}
On the other hand, the damping term in the exponential in Eq.\,(\ref{FinalCorrectCurvedProbability}) is given in terms of the proper length $L_p$ as follows,
\begin{equation}\label{InfiniteDamp}
\exp\left(-\frac{L^2}{L^2_{\rm coh}}\right)    \approx\exp\left\{-\frac{L_p^2}{L^2_{\rm coh}}\left(1-\frac{2GM}{L_p}\left[\frac{M_*r}{\sqrt{\pi}}\,_2F_2\left(\tfrac{1}{2},\tfrac{1}{2};\tfrac{3}{2},\tfrac{3}{2};-\frac{M_*^2r^2}{4}\right)\right]_{r_A}^{r_B}\right)\right\}.
\end{equation} 
Similar to what we found with the $\mathcal{R}^2$-gravity model, the infinite-derivative model induces a decrease in the damping factor. This can be seen from the second term of the expansion of the generalized hypergeometric function: $\,_2F_2(a,b;c,d;z)=1+\frac{ab}{cd}\frac{z}{1!}+\frac{a(a+1)b(b+1)}{c(c+1)d(d+1)}\frac{z^2}{2!}+\ldots$ \cite{MathHandBook}.  
\section{Conclusion \& discussion}\label{Conclusion}
We have studied neutrino flavor oscillations in general static and spherically symmetric curved spacetimes by treating each mass eigenstate of the linear superposition as a wave packet. We have distinguished between two different approaches for implementing the wave packet formalism. The first approach we examined consists in building the wave packet using the plane waves of the mass eigenstates after they reach the detector by first computing the accumulated quantum phase of each of those plane waves. The second approach we examined consists in building first the wave packet right from the neutrinos source before working out the effect of curved spacetime on such wave packets as they propagate toward the detector. We found that the two approaches are fundamentally different in that they lead to distinctly different results for the neutrino flavor transition probabilities in curved spacetimes. In addition, unlike the first approach, the results we found with the second approach naturally reduce to the flat-space results one obtains within the wave packet treatment in Minkowski spacetime. 

The important effect of curved spacetime on neutrino flavor oscillations implied by the second approach but missed by the first is the widening of the wave packets along their journey. The first approach does not entail such a widening whereas the second does in a way that neatly displays the role of the spacetime metric components. We exploited in this paper the explicit contribution of the metric components to both the enhancement of the flavor transition probability due to the widening of the wave packets and the damping of the transition probability due to the different group velocities the different mass eigenstates have. We examined various spacetime metrics that emerge from general relativity and which are relevant to astrophysics. We have examined the exterior Schwarzschild solution that describes the gravitational field outside massive objects and the interior Schwarzschild solution that describes the gravitational field inside spherically symmetric and homogeneous astrophysical objects. Although the latter is a rough approximation of the gravitational field inside real astrophysical objects, it has the merit of providing a neat effect of gravity on neutrino flavor oscillations as they travel inside matter.

Since our method works for any static and spherically symmetric metric, we have also applied it to the case of a spherical object embedded in an expanding de Sitter universe using the de Sitter-Schwarzschild metric. The effect of the cosmic expansion on neutrino flavor oscillations has been found to decrease the damping caused by the wave-packet nature of the neutrinos. 

Given the huge literature on attempts to go beyond general relativity by bringing modifications to the Einstein-Hilbert Lagrangian, we have also examined a few selected spacetime metrics emerging from such attempts. Our formulas showed a very distinct contribution to the flavor transition probability of higher-derivative terms added to the gravitational Lagrangian. Thus, our study did not only bring forward all the richness offered by taking neutrinos to be wave packets when considering the effect of curved spacetime on their flavor oscillations, but provided also a fresh look at the possible way of spotting gravitational effects beyond those provided by general relativity on neutrino flavor oscillations. Finally, one should note that the study we conducted here is entirely done within the quantum mechanical framework, but we expect that a field theoretical study, which will be attempted in a future work, will not modify much the conclusions we arrive at here.

\section*{Acknowledgments}
This work was supported by the Natural Sciences and Engineering Research Council of Canada (NSERC) Discovery Grant No. RGPIN-2017-05388; and by the Fonds de Recherche du Québec - Nature et Technologies (FRQNT). PS acknowledges support from Bishop's University via the Graduate Entrance Scholarship award.

\end{document}